\documentclass[10pt,letterpaper]{article}
\usepackage{opex3}
\usepackage{color}
\usepackage{amsmath,amsfonts}
\usepackage{subfigure}

\newcommand{\ket}[1]{\ensuremath{|#1\rangle}}

\begin{document}

\title{Iterative tailoring of optical quantum states with homodyne measurements}

\author{Jean Etesse$^{1,*}$, Bhaskar Kanseri$^1$ and Rosa Tualle-Brouri$^{1,2}$}
%\author{The Team$^{1}$}
\address{$^1$Laboratoire Charles Fabry, Institut d'Optique, CNRS, Universit\'e Paris Sud 11, \\ 2 avenue Augustin Fresnel, 91127 Palaiseau cedex, France\\
$^2$Institut Universitaire de France, 103 boulevard Saint-Michel, 75005 Paris, France}

\email{$^*$jean.etesse@institutoptique.fr} %% email address is required



\begin{abstract}
As they can travel long distances, free space optical quantum states are good candidates for carrying information in quantum information technology protocols. These states, however, are often complex to produce and require protocols whose success probability drops quickly with an increase of the mean photon number. Here we propose a new protocol for the generation and growth of arbitrary states, based on one by one coherent adjunctions of the simple state superposition $\alpha\ket{0}+\beta\ket{1}$. Due to the nature of the protocol, that allows for the use of quantum memories, it can outperform existing protocols.
\end{abstract}

 % REPLACE WITH CORRECT OCIS CODES FOR YOUR ARTICLE, MINIMUM OF TWO; Avoid using the OCIS codes for "General" or "General science" whenever possible.
%For a complete list of OCIS codes, visit: http://www.opticsinfobase.org/submit/ocis/

%%%%%%%%%%%%%%%%%%%%%%% References %%%%%%%%%%%%%%%%%%%%%%%%%

%%%%%%%%%%%%%%%%%%%%%%%%%%  body  %%%%%%%%%%%%%%%%%%%%%%%%%%
\section{Introduction}
Engineering of arbitrary mesoscopic quantum states of light is a challenging task. Impressive results were already obtained using giant enhancement in superconducting cavities  \cite{deleglise08,Vlaktakis13}, and protocols were proposed to generate arbitrary states with such systems \cite{vogel93}, but the trapped state cannot be used for quantum communication protocols. In the case of free space propagating quantum states of light, the most common method for optical state engineering is to generate the state directly, by using two entangled beams and by performing a measurement on one of these, either by click counting \cite{bimbard10,yukawa13} or by homodyning \cite{Jeong06, Babichev04, Laghaout13}. Schr\"odinger cat states of light for instance, consisting in a coherent superposition of two coherent sates and composing a basic resource for quantum information processing, have been produced using these techniques \cite{Ourjoumtsev08,Gerrits10}.\\

Building a state step by step is however necessary in order to grow its size, as the above mentioned methods are highly inefficient for large output states. Some protocols based on photon addition \cite{Dakna99} or subtraction \cite{Fiurasek05} propose this iteration of operations, but with the use of photon detection events which imply very low success probability. On the other hand, iterative generation based on homodyning has proven to be very efficient \cite{Etesse14}.\\
We propose here a generalization of the results presented in \cite{Etesse14}: we present a setup for the generation and for the growth of arbitrary quantum states of light by the successive application of a simple protocol that will be described and explicitly calculated in a particular case in section \ref{Protoc}, and whose performances will be discussed in section \ref{secperf}.

\section{Protocol}
\label{Protoc}
The idea of the protocol is to build a superposition containing up to $n+m$ photons by the ``mixing" of two superpositions containing up to $n$ and $m$ photons. Let us first see the simple case where $n=m=1$.
\subsection{Simple case}
\label{secelem}
The resource that we need to feed our protocol is the elementary superposition of vacuum and a single photon :
\begin{equation}
|\psi^{(1)}\rangle=\alpha|0\rangle+\beta|1\rangle.
\label{resource}
\end{equation}
This superposition can be experimentally generated by using homodyne conditioning \cite{Babichev04} or photon counting \cite{Resch02}, and we will assume it to be available on demand. Let us first see how the mixing of two states of the form (\ref{resource}) on a beamsplitter can produce an arbitrary superposition with two photons. 

\begin{figure}[!h]
\begin{center}
\subfigure[]{
\includegraphics[width=6.5cm]{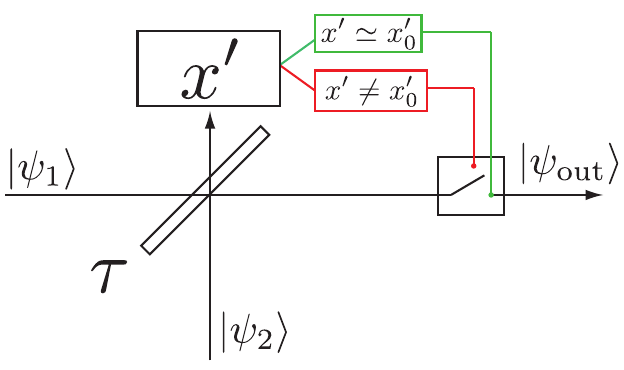}}
\subfigure[]{
\includegraphics[width=6.5cm]{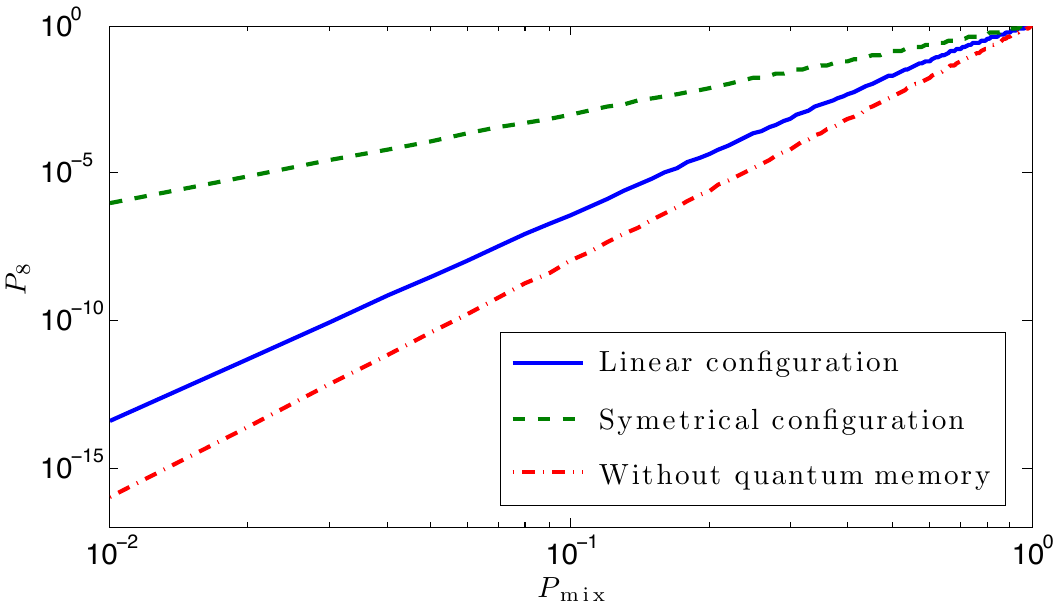}}
\caption{Protocol for generation and growth of arbitrary states (a) Elementary protocol: two states $\ket{\psi_1}$ and $\ket{\psi_2}$ are mixed on a beamsplitter with transmission $\tau$. A homodyne measurement is performed in one output arm, and the generation of a state $\ket{\psi_{\rm out}}$ in the other arm is conditioned upon the homodyne event $x'=x'_0$. (b) Total sucess probability of a protocol involving eight input resource states (\ref{resource}) in a linear (solid blue line) and a symetrical (dashed green line) configuration. The total success probability of a linear protocol not using any quantum memory is shown for comparison (dot-dashed red line) .}
\label{elemprotoc}
\end{center}
\end{figure}

The principle is shown on the figure \ref{elemprotoc} (a): two of these resource states $\ket{\psi_1}=|\psi^{(1)}_1\rangle=\alpha_1|0\rangle+\beta_1|1\rangle$ and $\ket{\psi_2}=|\psi^{(1)}_2\rangle=\alpha_2|0\rangle+\beta_2|1\rangle$ are sent on a beamsplitter with transmission $\tau$, and a homodyne detection is performed on one output arm. When the homodyne conditioning is successful ($x'=x'_0$), the wavefunction of the other output arm state is projected on:

\begin{equation}
\psi_{\rm out}(x)\propto (a_1 x-b_1)(a_2 x-b_2) e^{-\frac{x^2}{2}},
\label{superpos012gal}
\end{equation}
with $a_1 = \sqrt{2}\beta_1\tau$, $a_2=\sqrt{2}\beta_2\rho$, $b_1=-(\alpha_1+\sqrt{2}\beta_1\rho x_0')$, $b_2=\alpha_2+\sqrt{2}\beta_2\tau x_0'$ and $\rho=\sqrt{1-\tau^2}$.\\

Let us then remember the expression of the wavefunction of a Fock state $\langle x|k\rangle=1/(\pi^{1/4}\sqrt{2^kk!})H_k(x)exp(-x^2/2)$, where $H_k(x)$ is the $k^{th}$ Hermite polynomial (of degree $k$): this form tells us that any superposition with up to $n$ photons will have a wavefunction that will be written as a polynomial of degree up to $n$, times a gaussian of unit variance. This comes from the fact that the $H_k$ polynomials are a basis of $\mathbb{C}[X]$.\\

In the present case, the form (\ref{superpos012gal}) is the general writing of an arbitrary polynomial of degree up to two times a gaussian (all the polynomial can be splitted in $\mathbb{C}[X]$). According to the previous remark, this means that the corresponding state is an arbitrary superposition of up to two photons, whose parameters can be adjusted with $\alpha_i$, $\beta_i$, $\tau$ and $x'_0$.

\subsection{General case}
Let us generalize the idea of the previous paragraph by recurrence. Let us suppose that we have been able to generate a superposition with up to $n$ and $m$ photons, and see how we can generate a superposition with $n+m$ photons.\\
Mathematically speaking, it means that we assume to have two states $|\psi^{(n)}\rangle$ and $|\psi^{(m)}\rangle$ whose wavefunctions can be written as 
\begin{subequations}
\begin{eqnarray}
\psi^{(n)}(x)&\propto& P_{n}(x)e^{-\frac{x^2}2}\\
\psi^{(m)}(x)&\propto& P_{m}(x)e^{-\frac{x^2}2},
\end{eqnarray}
\end{subequations}
where $P_n$ (resp. $P_m$) is a polynomial of degree $n$ (resp. $m$).

%\begin{figure}[!h]
%\begin{center}
%\includegraphics[width=8cm]{figures/figure2v3}
%\caption{Protocol for the generation of a superposition with up to $n+m$ photons ($|\psi^{(n+m)}\rangle$) by the mixing of two superpositions of up to $n$ ($|\psi^{(n)}\rangle$) and $m$ ($|\psi^{(m)}\rangle$) photons.}
%\label{generprotoc}
%\end{center}
%\end{figure}

Let us mix these two states according to the same scheme of figure \ref{elemprotoc} (a), by feeding the setup with $\ket{\psi_1}=\ket{\psi^{(n)}}$ and $\ket{\psi_2}=\ket{\psi^{(m)}}$. The state that we will thereby generate can be written as :

\begin{equation}
\psi_{\rm out}(x)\propto P_n(\tau x+\rho x'_{0})P_m(\tau x'_{0}-\rho x) e^{-\frac{x^2}{2}}.
\label{superposgal}
\end{equation}
The wavefunction of this state is of the form of an arbitrary polynomial of degree $n+m$, times a gaussian of unit variance, and according to what was noticed previously, this state is then an arbitrary superposition of up to $n+m$ photons.\\

The protocol transformation is true for $n=m=1$ and can be iterated for any $n$ or $m$, which means that it is true for any $n$ and $m$: we have proven that the use of the simple protocol of figure \ref{elemprotoc} (a) iterated $n$ times and fed by superposition of the form (\ref{resource}) can generate arbitrary superpositions of up to $n$ photons.

\subsection{Structuration of the protocol}
A great advantage of our protocol is that it allows for the use of quantum memories between each homodyne conditioning. These devices are currently developing very quickly \cite{Yoshikawa13}, and they give a potential increase in the total sucess probability if the number of iterations increases. In \cite{Etesse14} is treated in detail the way one should design the protocol in order to maximize the total success probability: the idea is to perform a maximum of operations in parallel. Two types of configurations can then be distinguished: a linear configuration in which the output states of the protocol are mixed with a resource state iteratively, and a symetrized configuration in which the inputs of all the elementary protocols have been produced by using the same number of resource states. This configuration allows for the simultaneous realization of homodyne conditionings, contrary to the linear one.\\
 If one assumes for instance that all the homodyne conditioning have the same success probability $P_{mix}$, figure \ref{elemprotoc} (b) shows the tremendous increase in the total success probability allowed with the use of quantum memories in a symetrical protocol configuration, in the case where eight input resource states are used.

%\begin{figure}[!h]
%\begin{center}
%\includegraphics[width=10cm]{figures/probasucconfen}
%\caption{Total sucess probability of a protocol involving eight input resource states (\ref{resource}) in a linear (solid blue line) and a symetrical (dot-dashed red line) configuration. The total success probability of a linear protocol not using any quantum memory is shown for comparison (dashed green line).}
%\label{design}
%\end{center}
%\end{figure}

%citer les mŽmoires quantiques \cite{Yoshikawa13}

 \subsection{Example}
Let us see with a concrete example how the protocol can be used to generate states of light, by studying how the protocol can output an arbitrary superposition of the form proposed in \cite{Fiurasek05}:
\begin{equation}
|\psi_{\rm targ}\rangle=\frac{1}{\sqrt{1+|c_0|^2+|c_1|^2}}(c_0|0\rangle+c_1|1\rangle+|2\rangle).
\label{target}
\end{equation}
We have previously shown that we could generate any superposition of this kind by the use of equation (\ref{superpos012gal}): what should be the parameters of our protocol to generate the state (\ref{target})?\\

First, given the expression of the target state, the weight of the two photon is never 0, so we know that $a_1a_2\neq0$. The two roots of the polynomial of the wavefunction (\ref{superpos012gal}) are then $b_1/a_1$ and $b_2/a_2$. These should then be identified to the roots of the polynomial in the wavefunction (\ref{target}) :

\begin{equation}
\psi_{\rm targ}(x)\propto\Big[x^2+{c_1}x+\frac{c_0}{\sqrt{2}}-\frac12\Big]e^{-\frac{x^2}2}.
\end{equation}
To simplify the calculation, we are going to suppose that the discriminant of this equation is positive $\Delta=c_1^2-4(c_0/\sqrt{2}-1/2)>0$, then according to the expressions of $a_1$, $b_1$, $a_2$ and $b_2$, we find the results:
%\begin{subequations}
%\begin{eqnarray}
%\alpha_1&=&\epsilon_1\sqrt{\frac{2\big[x_1\tau-\rho x'_0\big]^2}{1+2\big[x_1\tau-\rho x'_0\big]^2}},\qquad\qquad\beta_1^2=1-|\alpha_1|^2,\\
%\alpha_2&=&\epsilon_2\sqrt{\frac{2\big[x_2\rho-\tau x'_0\big]^2}{1+2\big[x_2\rho-\tau x'_0\big]^2}},\qquad\qquad\beta_2^2=1-|\alpha_2|^2,
%\end{eqnarray}
%\end{subequations}

\begin{subequations}
\begin{eqnarray}
\beta_1&=&\epsilon_1\frac{1}{\sqrt{1+2\big[x_1\tau-\rho x'_0\big]^2}},\qquad\qquad\alpha_1=\sqrt{1-|\beta_1|^2},\\
\beta_2&=&\epsilon_2\frac{1}{\sqrt{1+2\big[x_2\rho-\tau x'_0\big]^2}},\qquad\qquad\alpha_2=\sqrt{1-|\beta_2|^2},
\end{eqnarray}
\end{subequations}
with $x_1=\frac{\sqrt{\Delta}+c_1}{2}$, $x_2=\frac{\sqrt{\Delta}-c_1}{2}$, $\epsilon_1=sign(x_1\tau-\rho x'_0)$ and $\epsilon_2=sign(x_2\rho-\tau x'_0)$.\\
%which are valid under the constraint $x'_0\leq\min\Big(\frac{\sqrt{\Delta}+c_1}{2}\frac{\tau}{\rho},\frac{\sqrt{\Delta}-c_1}{2}\frac{\rho}{\tau}\Big)$.\\
%For a numerical application, let us consider the case of the two photon Fock state generation. In this simple case, $c_0=c_1=0$, and the results can be rewritten as:
%\begin{subequations}
%\begin{eqnarray}
%\alpha_1^2&=&\frac{2\big[\frac{\tau}{\sqrt2}-\rho x'_0\big]^2}{1+2\big[\frac{\tau}{\sqrt2}-\rho x'_0\big]^2},\qquad\qquad\beta_1^2=1-|\alpha_1|^2,\\
%\alpha_2^2&=&\frac{2\big[\frac{\rho}{\sqrt2}-\tau x'_0\big]^2}{1+2\big[\frac{\rho}{\sqrt2}-\tau x'_0\big]^2},\qquad\qquad\beta_2^2=1-|\alpha_2|^2,
%\end{eqnarray}
%\label{coeffs2phot}
%\end{subequations}

For a numerical application, let us consider the case of the simple superposition $2^{-1/2}(\ket{1}+\ket{2})$. In this simple case, $c_0=0$ and $c_1=1$, and the previous results can be rewritten as:
%\begin{subequations}
%\begin{eqnarray}
%\alpha_1^2&=&\frac{2\big[\frac{\tau}{\sqrt2}-\rho x'_0\big]^2}{1+2\big[\frac{\tau}{\sqrt2}-\rho x'_0\big]^2},\qquad\qquad\beta_1^2=1-|\alpha_1|^2,\\
%\alpha_2^2&=&\frac{2\big[\frac{\rho}{\sqrt2}-\tau x'_0\big]^2}{1+2\big[\frac{\rho}{\sqrt2}-\tau x'_0\big]^2},\qquad\qquad\beta_2^2=1-|\alpha_2|^2,
%\end{eqnarray}
%\label{coeffs2phot}
%\end{subequations}
\begin{subequations}
\begin{eqnarray}
\beta_1&=&\epsilon_1\frac{1}{\sqrt{1+2\big[\frac{\sqrt{3}+1}{2}\tau-\rho x'_0\big]^2}},\qquad\qquad\alpha_1=\sqrt{1-|\beta_1|^2},\\
\beta_2&=&\epsilon_2\frac{1}{\sqrt{1+2\big[\frac{\sqrt{3}-1}{2}\rho-\tau x'_0\big]^2}},\qquad\qquad\alpha_2=\sqrt{1-|\beta_2|^2},
\end{eqnarray}
\label{Applnum}
\end{subequations}
with $\epsilon_1=sign(\frac{\sqrt{3}+1}{2}\tau-\rho x'_0)$ and $\epsilon_2=sign(\frac{\sqrt{3}-1}{2}\rho-\tau x'_0)$.\\
We clearly see that we have two supplementary degrees of freedom for the generation of our state: $x'_0$ and $\tau$. As they can be freely adjusted, they will give us the possibility to maximize the success probability of the operation. This is what we are going to study in the next section.

\section{Performances}
\label{secperf}
\subsection{Success probability}
Obviously, heralding on events matching exactly the homodyne condition $x'=x'_0$ will lead to a zero success probability, so one has to accept the events within a window $x'\in[x'_0-\Delta x,x'_0+\Delta x]$. Increasing its width $\Delta x$ will increase the success probability, but at the cost of a decrease in the quality of the state. In order to perform the study of the success probability of the protocol we propose to fix a target fidelity of the state we want to achieve, and to optimize the heralding width of the homodyne conditioning in order to maximize the success probability of the operation.\\

\begin{figure}[!h]
\begin{center}
\subfigure[]{
\includegraphics[width=5.8cm]{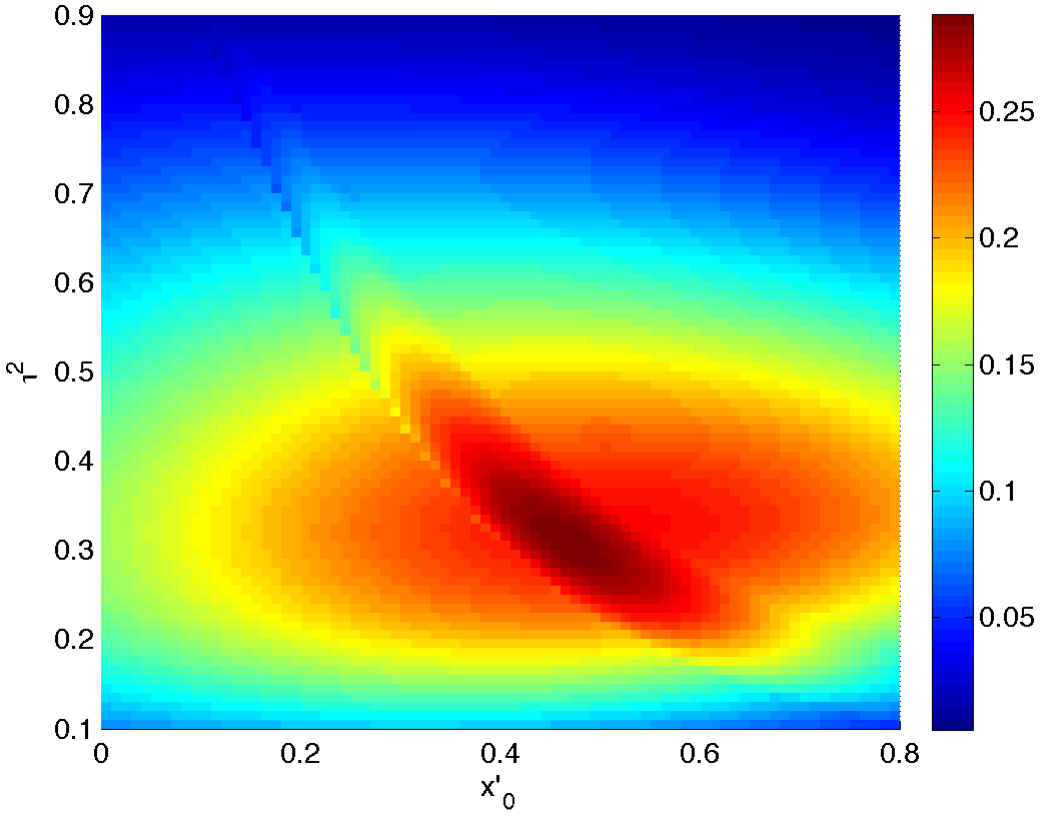}}
\subfigure[]{
\includegraphics[width=7.2cm]{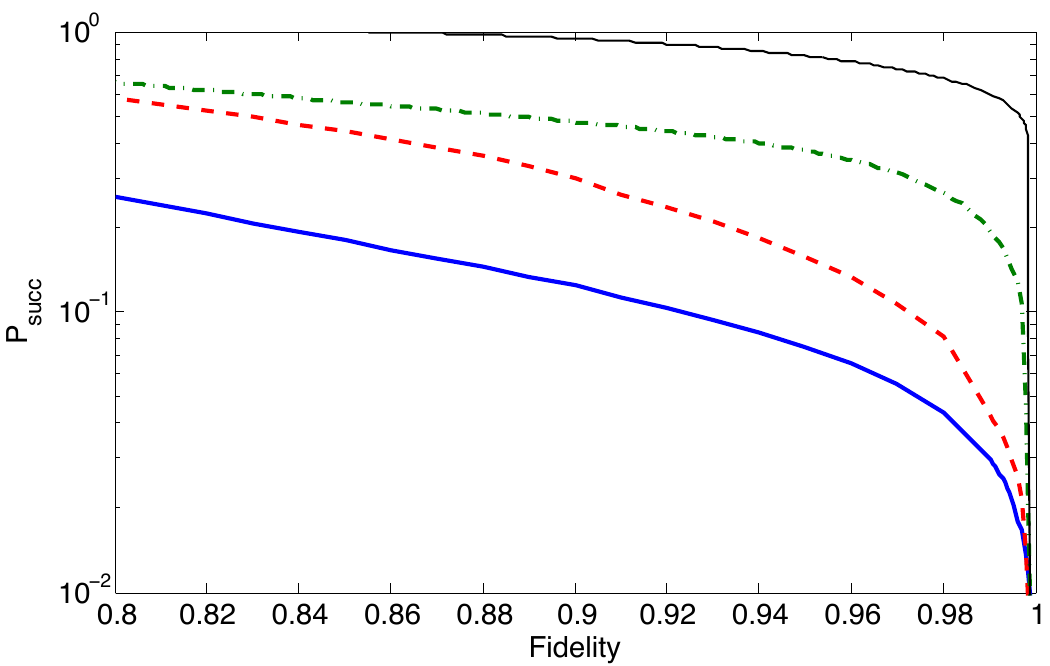}}
\caption{Optimization of the success probability. (a) Success probability of the protocol for the generation of the state $2^{-1/2}(\ket{1}+\ket{2})$ as a function of the quadrature conditioning $x'_0$ and the energy transmission of the beamsplitter $\tau^2$. The coefficients of the resource states are given by equation (\ref{Applnum}), and the target fidelity is 90\%. (b) Optimized success probability for the states (\ref{psi1}) (solid blue), (\ref{psi2}) (dashed red), (\ref{psi3}) (dot-dashed green) and (\ref{psi4}) (solid thin black).}
\label{optimparam}
\end{center}
\end{figure}

Let us first focus on the previous example: the state superposition $2^{-1/2}(\ket{1}+\ket{2})$. We have seen that the coefficients $\tau$ and $x'_0$ could be freely chosen in order to generate it. By using equations (\ref{Applnum}), we can plot the success probability as a function of these two coefficients. Figure \ref{optimparam} (a) shows this for a target fidelity of 90\%, revealing that there is actually an optimal point for $(\tau^2,x'_0)$ around $(0.32,0.46)$ for the generation of the state, leading to almost 30\% success probability of generation.\\

This optimization can be performed on various states, showing some difference in the efficiency of production. For instance, figure \ref{optimparam} (b) shows the optimized success probability as a function of the target fidelity for the four states of the form (\ref{target}):
\begin{eqnarray}
\ket{\psi_1}&=&\ket{2}\label{psi1}\\
\ket{\psi_2}&=&\frac{1}{\sqrt2}(\ket{1}+\ket{2})\label{psi2}\\
\ket{\psi_3}&=&\frac{1}{\sqrt2}(\ket{0}+\ket{2})\label{psi3}\\
\ket{\psi_4}&=&\frac{1}{\sqrt3}(\ket{0}+\ket{1}+\ket{2})\label{psi4},
\end{eqnarray}
the optimization being performed on all the parameters that we can adjust for the state ($\alpha_i$, $\beta_i$, $\tau$ and $x'_0$).\\
We see that the success probability of our protocol is very high compared to other previously proposed setups. Indeed, the four states (\ref{psi1})-(\ref{psi4}) were also studied in \cite{Fiurasek05}, and provided success probabilities of the order of $10^{-5}$ for target fidelities of 90\%. In our case, these success probabilities are greater than 10\% and reach almost 100\% for the state (\ref{psi4}): this impressive behaviour is simply explained by the fact that the unconditioned state (100\% success probability by definition) has already 87\% fidelity with the target state.

\subsection{Imperfections}
Let us now study the influence of imperfections on the protocol. These will be taken of two different types: either from the resource state itself or from the homodyne detection used for the heralding.\\

For the sake of simplicity, and to picture precisely the effect of the protocol, we will consider in this paragraph the generation of states with a protocol fed by two single photon Fock states ($\alpha=0$ in (\ref{resource})). The Hong Ou Mandel effect \cite{Hong87} makes the one photon contribution vanish ($c_1=0$ in (\ref{target})), and we are left with superpositions of the kind $|\psi\rangle=\frac{1}{\sqrt{1+|c'_0|^2}}(c'_0|0\rangle-|2\rangle)$ with $c'_0\geq0$. We will focus on three particular cases: $c'_0=0$ (the two photon Fock state), $c'_0=1$ (equally weighted states) and $c'_0=1/\sqrt{2}$ (states created in \cite{Etesse14}). The generation of these states can be performed for instance with a symetrical beamsplitter and a conditioning on $x'_0=\frac{1}{\sqrt{2}}$, $x'_0=\sqrt{\frac{1+\sqrt{2}}{2}}$ and $x'_0=0$ respectively. We won't try to optimize the success probability here, as we want to estimate the effects of the imperfections only. The conditioning width $\Delta x$ will then be taken arbitrarily small.

\subsubsection{Imperfections of the single photons}
\begin{figure}[!h]
\begin{center}
\subfigure[]{
\includegraphics[width=6.43cm]{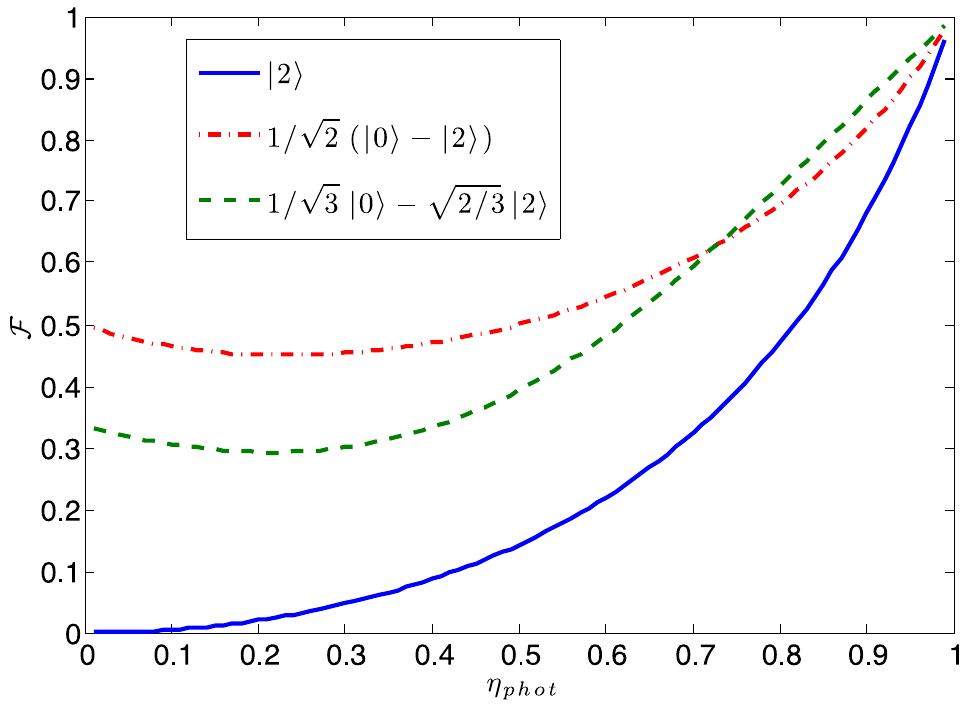}}
\subfigure[]{
\includegraphics[width=6.5cm]{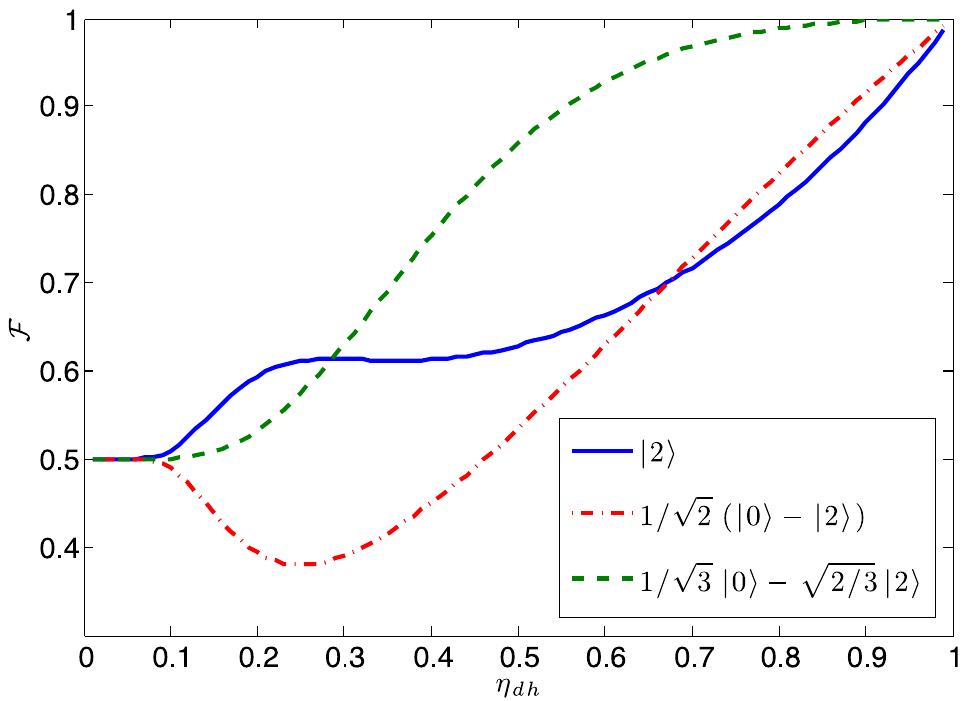}}
\caption{Influence of the imperfections on the quality of the output state in the case $c'_0=0$ (solid blue line), $c'_0=1$ (red dot-dashed line) and $c'_0=1/\sqrt{2}$ (dashed green line), for imperfections (a) on the photons or (b) on the homodyne detections.}
\label{fidelimperfphot}
\end{center}
\end{figure}

In the case where the photons used to feed the setup are imperfect, the performances of the protocol are deteriorated. The imperfections that we are going to take into account are the most common ones: the photons are no longer pure, but consist in a mixture of the single photon $|1\rangle\langle1|$ with vacuum  $|0\rangle\langle0|$. The respective weights give the quality of the state:
\begin{equation}
|\psi\rangle=\eta_{phot}|1\rangle\langle1|+(1-\eta_{phot})|0\rangle\langle0|.
\end{equation} 
The fidelity $\mathcal{F}$ of the output state as a function of the quality $\eta_{phot}$ of the photon is given in figure \ref{fidelimperfphot} (a). Obviously, when this quality tends to 1 (no admixture of vacuum), the fidelity of the output state tends to 1. In the opposite case, the fidelity tends to the square of the weight of vacuum in the superposition, as expected (the input state is only composed of vacuum when $\eta_{phot}=0$).\\
An interersting point is that the quality does not deteriorate faster as the weight of the two photon Fock state increases in the target state: the state with $c'_0=1$ deteriorates faster than the one with $c'_0=1/\sqrt{2}$ for input qualities $\eta_{phot}>0.7$, even if it has a smaller two photon component.\\
%the green curve shrinks faster than the red one for input qualities $\eta_{phot}>0.7$.\\
This figure shows that the quality of the input photons is a key property for the proper realization of the protocol, as the output fidelity is strongly dependant on it. For instance, an input fidelity of 90\% of the input single photons leads to an output fidelity of 70\% for the two photons Fock state output ($c'_0=0$).

\subsubsection{Imperfections of the homodyne detections}
Other imperfections that can be taken into account are the detection inefficiencies of the homodyne detections used for the conditioning. These imperfections (treated in detail in \cite{Etesse14}) can be shown to artificially increase the heralding width, and their effects are shown on figure \ref{fidelimperfphot} (b), for the same three cases mentioned previously.\\
Their effects are quite different, as the fidelity can remain reasonably high for low detection efficiencies. For instance in the case $c'_0=1/\sqrt{2}$, the output fidelity is above 90\% for detection efficiencies as low as 55\%, revealing a certain robustness of the protocol against its own imperfections. 

\section{Conclusion}
We have proposed a new protocol which enables the generation of arbitrary superpositions of a given number of photons, by the iterative use of a simple protocol based on a mixing on a beamsplitter followed by a homodyne conditioning measurement. This protocol is a real breakthrough regarding quantum engineering of states, as it consists in the building piece by piece of the state, and allows for the use of quantum memories in order to improve the success probabilities which can then be very high, on the contrary to all the previous protocols. Another great advantage of the homodyne conditioning technique is that it is very robust against detection inefficiencies.\\
 With all the recent advances in quantum memories technologies \cite{Yoshikawa13} as well as monomode single photon generation \cite{Morin12}, we believe that this proposal will open new perspectives in the field of quantum optical states engineering.

\section*{Aknowledgments}
We acknowledge support from the EU project ANR ERA-Net CHISTERA HIPERCOM.

\end{document}